\pdfoutput=1
\documentclass[12pt]{iopart}
\usepackage{iopams}
\usepackage{graphicx}
\usepackage{color}

\begin{document}

\title{Anomalous zipping dynamics and forced polymer translocation}

\author{A. Ferrantini, E. Carlon}

\address{Institute for Theoretical Physics, K.U.Leuven, Celestijnenlaan 200D, B-3001 Leuven, Belgium}
\ead{alessandro@itf.fys.kuleuven.be, enrico.carlon@fys.kuleuven.be}

\begin{abstract}
We investigate by Monte Carlo simulations the zipping and unzipping dynamics of two polymers connected by one end and subject to an attractive interaction between complementary monomers. In zipping, the polymers are quenched from a high temperature equilibrium configuration to a low temperature state, so that the two strands zip up by closing up a ``Y"-fork. In unzipping, the polymers are brought from a low temperature double stranded configuration to high temperatures, so that the two strands separate. Simulations show that the unzipping time, $\tau_u$, scales as a function of the polymer length as $\tau_u \sim L$, while the zipping is characterized by anomalous dynamics $\tau_z \sim L^\alpha$ with $\alpha = 1.37(2)$. This exponent is in good agreement with simulation results and theoretical predictions for the scaling of the translocation time of a forced polymer passing through a narrow pore. We find that the exponent $\alpha$ is robust against variations of parameters and temperature, whereas the scaling of $\tau_z$ as a function of the driving force shows the existence of two different regimes: the weak forcing ($\tau_z \sim 1/F$) and strong forcing ($\tau_z$ independent of $F$) regimes. The crossover region is possibly characterized by a non-trivial scaling in $F$, matching the prediction of recent theories of polymer translocation. Although the geometrical setup is different, zipping and translocation share thus the same type of anomalous dynamics. Systems where this dynamics could be experimentally investigated are DNA (or RNA) hairpins: our results imply an anomalous dynamics for the hairpins closing times, but not for the opening times.
\end{abstract}

\submitto{JSTAT}

\maketitle

\section{Introduction}

The aim of this paper is to study the zipping and unzipping 
dynamics of polymers.
Zipping occurs when two polymer strands, with attractive interactions
between complementary monomers, bind to form a double-stranded
conformation. This is the behavior of complementary DNA strands
forming a double helical structure by closing up a Y-fork in which
two single strands join into a double stranded segment (hence the name
zipping). The reverse transition, the unzipping, is the separation of
the two strands at high temperatures, which can also occur under the
effect of a mechanical force pulling the edges of a polymer. Mechanical
unzipping has been the subject of several studies in the past
\cite{lube00,bhat00,seba00,mare01,math06,kuma10}, due to its relevance to single
molecule experiments (see e.g. \cite{esse97,cocc01}). Equilibrium
properties of zipping transitions have also been investigated
\cite{kuma95,baie01,baie02,leon03}. We restrict ourselves here to the case of
zipping induced by attractive interactions between monomers, in absence
of mechanical forces. We show that the dynamics is anomalous and that it
is characterized by an exponent in agreement with that found in polymer
translocation \cite{vock08}.

\begin{figure}[ht!]
\includegraphics[height=6cm]{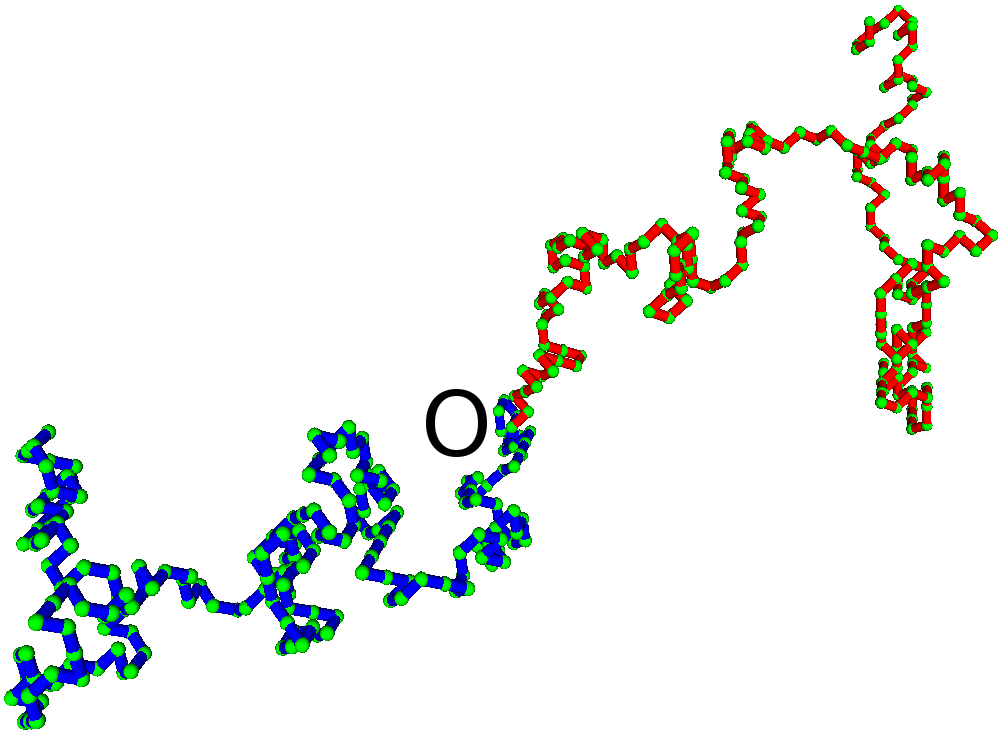}
\includegraphics[height=5.0cm]{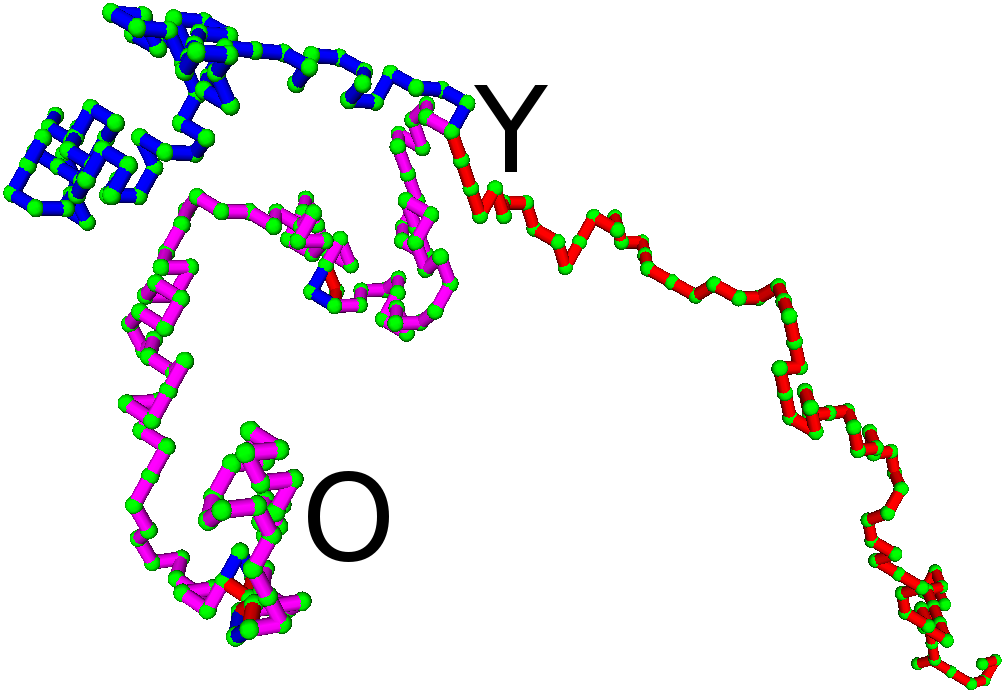}
\caption{Snapshots of two different polymer configurations (model~B,
$L=193$, $\omega=0.02$). Left panel: an inifinite temperature
equilibrium configuration. Right panel: an intermediate configuration
during zipping. ``O'' denotes the joint end of the two polymers, while
``Y'' is the branching point where the double-stranded stretch joins
the single strands.}
\label{zipconf}
\end{figure}

In the simulations two polymer strands are attached to each other
from one end and are prepared in a high temperature equilibrium state
(see Fig.~\ref{zipconf}, left panel). The system is then quenched to
low temperatures, below the thermal unzipping temperature, so that a
double-stranded conformation gets formed in the course of time.
We consider lattice polymers undergoing Monte Carlo dynamics with
local flip moves (see next section for details) which do not violate the
self- and mutual avoidance between the strands. This corresponds to a
Rouse dynamics, while hydrodynamics effects are neglected. In addition,
zipping here
occurs without the winding of the strands around each other, as in
DNA molecules, but by a pairing of the two strands. The right panel of
Figure~\ref{zipconf} shows an example of an intermediate conformation
during the zipping process.  The branch connecting the points ``O"
and ``Y" is double stranded, and ``Y" is the contact point between the
zipped part and the single stranded ends. We will also briefly discuss
the reverse case of unzipping dynamics where an equilibrated double
stranded polymer is brought at high temperatures. This case turns out to
be less interesting because the dynamics is not anomalous. However, it
provides some insights on the origin of the anomalous dynamics observed
during zipping.

Our main interest is the scaling of the zipping time $\tau_z$ (to be
defined more precisely later) as a function of the length of the strands
$L$. Using simulations we find that $\tau_z \sim L^{1.37(2)}$. This
exponent is in good agreement with that found in the study of a polymer
performing a biased translocation through a narrow pore \cite{vock08}. The
same type of scaling has been recently observed in the case of a related
problem of polymer adsorption on a flat substrate \cite{panj09}.

Several publications have appeared in the recent
literature about translocation of polymers through a pore
\cite{kant04,huop07,vock08,gaut08,bhat10}. The pore is a small aperture on
a plane which is sufficiently narrow to allow the passage of one or a few
monomers at a time. The time needed by the polymer to cross the pore is
referred to as translocation time ($\tau_t$). Although there is a general
consensus about the fact that the translocation time scales as a power
of the polymer length, the precise numerical value of the exponent has
been the subject of some debate~\cite{kant04,huop07,vock08,gaut08,bhat09,bhat10}.

In Ref.~\cite{kant04}, then taken up in Ref.~\cite{cacc06}, a lower
bound limit $\tau_t \geq L^{1+\nu}$ was
derived for the scaling of the translocation time. Here $\nu$ is the Flory
exponent ($\nu = 0.588$). This bound was obtained (under some simplifying
assumptions) by comparison with the motion of a driven polymer in the
bulk, i.e. in absence of the separating plane. Early numerical simulation
of the translocation times \cite{huop06} found a scaling of the type
$\tau_t \sim L^{1.59}$, particularly in the limit of high pore friction.
More recently, Vocks {\it et al.} \cite{vock08} derived the following
scaling for the translocation time $\tau_t \sim L^\alpha$ with $\alpha =
(1+2\nu)/(1+\nu)$, yielding thus $\alpha = 1.37$.  This prediction is
based on memory effects which arise due to an imbalance in the chain
tension in the vicinity of the pore and are observed also in the case of
unbiased translocation \cite{panj07}. Numerical simulations~\cite{vock08}
confirm the theoretical prediction for $\alpha$.  
Recent simulations of  forced translocation \cite{luo09} also found
$\alpha \approx 1.37$ in a regime referred to as fast translocation.

The adsorption of polymeric chains onto a surface has also been subject
of experimental \cite{doug97}, as well as computational and theoretical
studies \cite{pono00,desc06,bhat08,panj09}.  In the strong adsorption
regime, the characteristic adsorption time scales with the chain length
as $\tau_a \sim L^{1 + \nu}$ \cite{pono00,desc06}.  The exponent $1+\nu$
was explained with a two-phase model in which an intermediate adsorbed
configuration is viewed as consisting of two non-equilibrium phases: the
{\it adsorbate} and the non-adsorbed phase (the {\it corona}), connected
by a stretched part of the chain (the {\it stem}). More recent simulations
\cite{panj09} showed two regimes for polymer adsorption, depending on
the adsorption energies: $\tau_a \sim L^{1.37}$ for weak adsorption
energies and $\tau_a \sim L^{1.59}$ for strong adsorption energies.

In view of these results, and motivated by the recent interest in
anomalous dynamics in translocation and adsorption, we investigate here
the possibility of observing these effects in the context of polymers
zipping. We find indeed anomalous dynamics in the zipping process, but not
in the reverse case of unzipping. In the zipping case we investigate the
effects of changing temperature and other parameters in the system. Our
results show that there is only one regime in this system with a scaling
of the zipping time of the type $\tau_z \sim L^{1.37}$. Instead, we do
not observe a regime with $\tau_z  \sim L^{1.59}$, as seen in forced
translocation \cite{luo09} and in adsorption \cite{panj09}.

\section{Model}

The model discussed here was also used in a recent study of renaturation
dynamics \cite{ferr10}, where the scaling properties of the nucleation
rates of complementary polymers were investigated. We consider two
polymers defined on a face-centered-cubic lattice and joined by one
end. We label the monomers of the two strands starting from the common
monomer (corresponding to $i=0$), with $i=1,2,\ldots L$. The two strands
are self- and mutually avoiding, with the exception of monomers with the
same index $i$, which are referred to as complementary monomers. Two
complementary monomers can indeed bind by overlapping on the same
lattice site. 

The polymers undergo Rouse dynamics which consists of local corner-flip
or end-flip moves that do not violate self- and mutual avoidance. The
overlap between complementary monomers, which thus form a bound pair,
is always accepted as a move. The opposite move, that of detaching two
bound complementary monomers, is accepted with probability $\omega = \exp
( -{\varepsilon}/{k_B T} )$, so that detailed balance is satisfied. Here
$\varepsilon>0$ is the binding energy, $k_B$ the Boltzmann constant and
$T$ the temperature. An elementary move consists in selecting a random
monomer on one of the two strands. If the selected monomer is unbound a
local flip move is attempted. If the selected monomer is a bound monomer
there are two possibilities. Either a local flip of the chosen monomer
is attempted, and if accepted, this move results in the bond breakage;
or a flip move of both bound monomers is generated, which does not break
the bond between them. We fix the rate of single monomer move to $1$
and choose a rate $p_d$ for the double strand move. In most of the
computations we took $p_d=1$, but we also considered different values
of $p_d$, in particular $p_d < 1$, which implies a reduced ``mobility"
for the double strands. The limiting value of $p_d=0$ corresponds to a
double strand dynamics which can evolve only through bond breakage and
which is very unlikely to happen at low temperatures. In this case the
double stranded configuration remains basically ``frozen" during the
zipping process.

We will consider two models of zipping, as done in the study of mechanical
unzipping of polymers~\cite{mare01}. We will refer to them as model~B and model~Y.
In model~B one has plain zipping with no constraints and the formation of bubbles
(loops) is permitted. In model~Y all loops are suppressed. In this case we impose
the constraint that monomer $i$ can bind to its complementary only if monomer
$i−1$ is already bound. In this case zipping
proceeds strictly sequentially from the joining point of the two polymers
to the opposite ends of the two strands. With the flip moves used
in the model two strands or part of the same strand never cross. There is
however a possibility that the two strands ``cut through'' each other as
follows. Two monomers on different strands with the same index $i$ bind to
each other (by overlapping on the same lattice site) and then unbind in a
different direction from the original one, which may result in a strand
crossing. One can easily realize that this crossing may happen only in
model~B. In addition unbinding is rare at very low temperature, so the
crossing may be relevant only close to the unbinding temperature.  As in
our analysis we do not observe differences in the scaling of the zipping
times in the two models we conclude that the crossing through binding
and unbinding at homologous sites is irrelevant for the universal 
scaling behavior of the dynamics.

\begin{figure}[t]
\centering
\includegraphics[height=8.0cm]{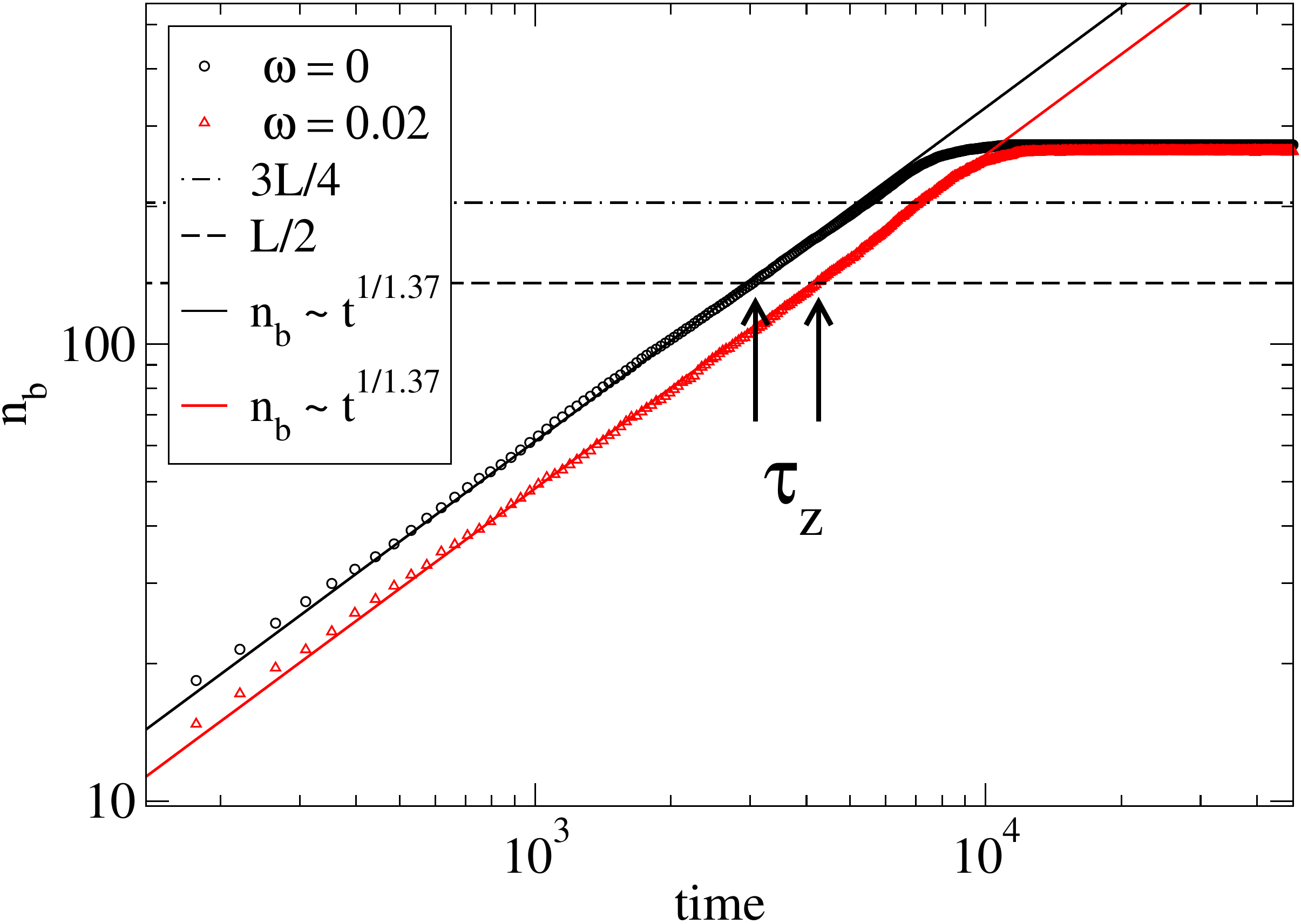}
\caption{Number of bound monomers $n_b(t)$ as a function of time $t$
(measured as the number of attempted Monte Carlo moves per monomer). The
length of the chains is $L=273$ and the dynamics is that denoted as model~B.
The two sets of data refer to two different temperatures: $\omega=0$
($T=0$) and $\omega=0.02$ ($T \approx T_c/2$). The straight line has
slope $\displaystyle 1/1.37$ and approximates well the observed power-law
increase of $n_b(t)$. The arrows denote the values of $\tau_z$ obtained from $n_b(\tau_z, L)
=L/2$. Using as an alternative criterion $n_b(\tau_z, L) = 3L/4$
(dotted-dashed line) would produce different estimates for $\tau_z$,
which however scale with the same power-law behavior as for $L/2$
(see text).
}
\label{tnsn}
\end{figure}

\section{Results}

In the simulations we monitored the number of bound bases $n_b(t)$ as
a function of time $t$. Figure~\ref{tnsn} shows a plot of $n_b(t)$ for
two different temperatures for $L=273$ and averaged over $50$ different
realizations. In both cases the number of bound monomers increases in time
until a saturation value, corresponding to the zipping of the polymer over
its entire length, is reached. For the scaling of $n_b$ we expect:
\begin{equation}
n_b (t,L) = L f(t/L^\alpha)
\label{scal_nbtL}
\end{equation}
with $f()$ a scaling function. For large values of $x=t/L^\alpha$,
$n_b$
reaches a saturation value $n_b \sim L$, hence $f(x) \to 1$ for $x \gg
1$. For $x \ll 1$ (short times) $n_b$ should be independent of $L$,
hence
\begin{equation}
n_b (t,L) \sim t^{1/\alpha}
\end{equation}
One can thus extract the exponent governing $\alpha$ from a power-law fit
of $n_b$ vs. $t$ in for very long polymers, to avoid finite size effects.
The solid line in Fig.~\ref{tnsn} shows that the data for $L=273$ are
consistent with a value $\alpha \approx 1.37$. However, due to the
presence of a saturation value one has to limit the analysis of $n_b(t)$
to some maximal time $t_{\rm max}$. In addition, the short time behavior
is usually affected by deviations from the asymptotic regime (one notices
some curvature of the data at early times in Fig.~\ref{tnsn}). Hence the
analysis of the slope of $\log n_b$ vs. $\log t$ can only be performed
on a limited time interval, which introduces some arbitrariness in the
procedure and uncontrolled errors on the value of $\alpha$.

\begin{figure}[ht!]
\centering
\includegraphics[height=8cm]{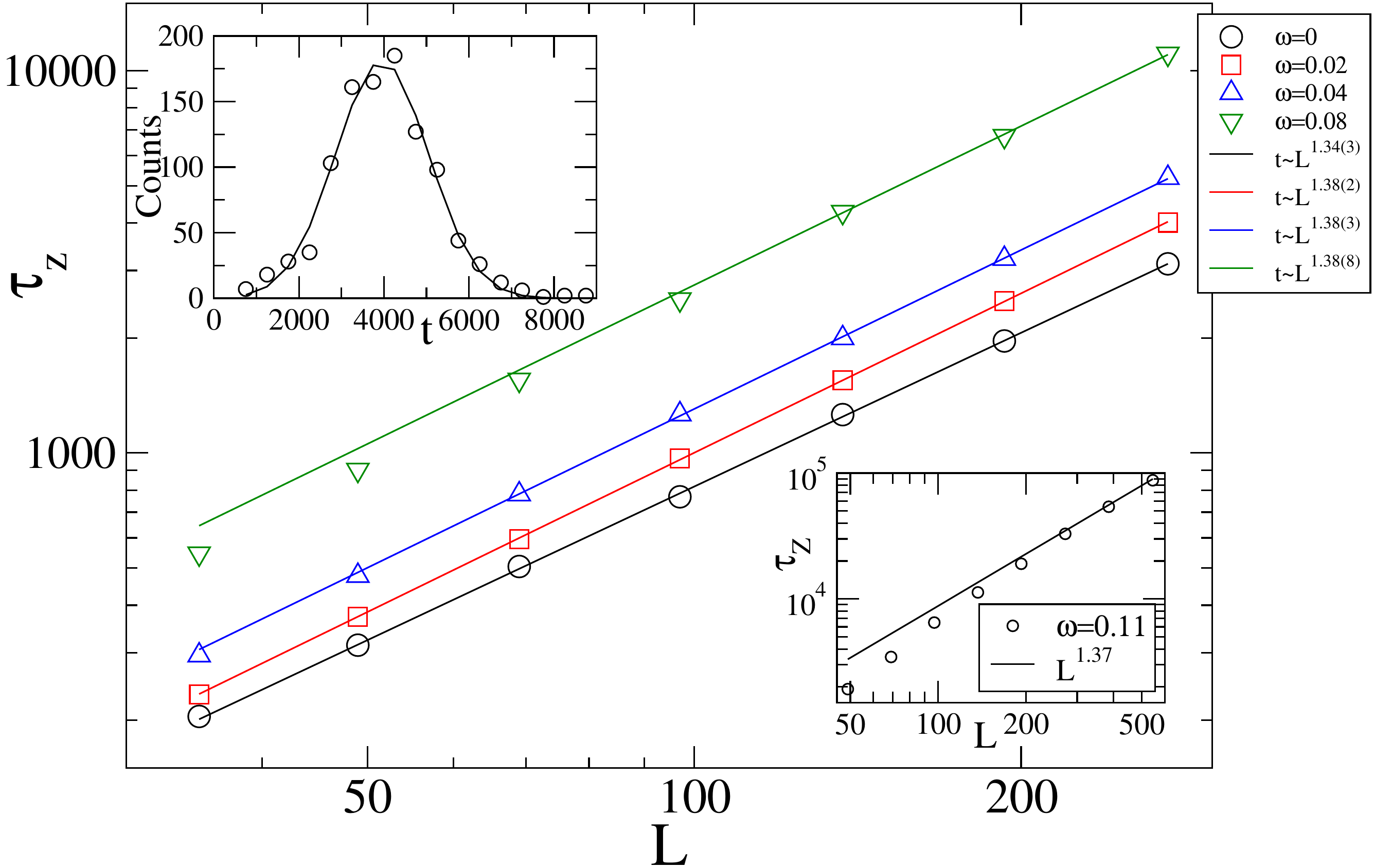}
\caption{Log-log plots of the zipping times (model~B) as a function of
the chain length for $\omega=0$ ($T=0$ or very strong bonds) in black,
$\omega=0.02$ ($T \simeq 0.5 T_c$) in red, $\displaystyle \omega=0.04$
($T \simeq 0.67 T_c$) in blue and $\displaystyle \omega=0.08$ ($T \simeq
0.8 T_c$) in green. In the last case the fit is on the last 3 points.
Bottom inset: $\tau_z$ vs. $L$ plot for $T \approx 0.94 T_c$
up to $L=800$.  The small $L$ data are characterized by a pronounced
curvature. For sufficiently long chain the asymptotic $\sim L^{1.37}$
scaling seems to be recovered.  Top inset: histogram of the zipping
times for $\omega=0.02$, $L=273$ obtained from $10^3$ independent
runs. The distribution of times is well-approximated by a gaussian, 
as shown by the fitting curve
(correlation coefficient $r=0.991$). The error on the mean is calculated
as $\displaystyle\delta\tau_z=\frac{\sigma}{\sqrt{N_r}} \simeq 35$,
with $\sigma^2$ being the variance of the gaussian and $N_r=1000$ being
the number of runs. This example shows that in this and the following
graphs error bars are smaller than symbol sizes.}
\label{rdm0.5}
\end{figure}

We used here a different approach and defined the zipping time as the
time needed to reach (for the first time) a configuration where half
of the monomers are bound. Figure~\ref{tnsn} illustrates how $\tau_z$
is obtained from the data for the two temperatures shown. In practice,
the simulations are stopped each time the number of bound monomers
reaches $L/2$, which also avoids long runs. To get an accurate estimate
of $\tau_z$ we averaged over about $10^3$ independent simulations.
Statistical errors are obtained from the standard deviations on these
different runs.

In Fig.~\ref{rdm0.5} we present a log-log plot of the zipping times as
a function of the strand length in some illustrative cases in model~B.
We note that the asymptotic scaling sets in already at relatively short
chains ($L \simeq 70$).  The data refer to different temperature values,
ranging from $T=0$ to $T \approx 0.8 T_c$ (we estimated the critical
value using equilibrium simulations with the pruned-enriched Rosenbluth
method \cite{gare90,gras97} to be $\omega_c \simeq 0.1266$).  The data
are in agreement with a constant value of $\alpha$ which we summarize as
$\alpha = 1.38(3)$.  At higher temperatures, close to the critical point
(see inset, which shows $\tau_z$ for $T \approx 0.94 T_c$ for lengths
up to $L=800$), the data show some curvature in the log-log scale,
but for sufficiently long polymers they seem to approach the same exponent
found in the low temperature cases (solid line). A recent paper on
forced translocation \cite{luo09} presented simulation evidences of the
existence of two different regimes in the scaling of the translocation
time as a function of $L$. At strong forcing a scaling would be governed
by an exponent $\alpha \approx 1.37$, while at weak forcing by $\alpha
\approx 1 + \nu \approx 1.58$~\cite{luo09}.  In the present model we do not see
a clear evidence of a second regime. However, close to $T_c$ the data
approach the asymptotic scaling $\sim L^{1.37}$ from ``below". Finite $L$
data are characterized by a higher running exponent.


\begin{figure}[ht!]
\centering
\includegraphics[height=8cm]{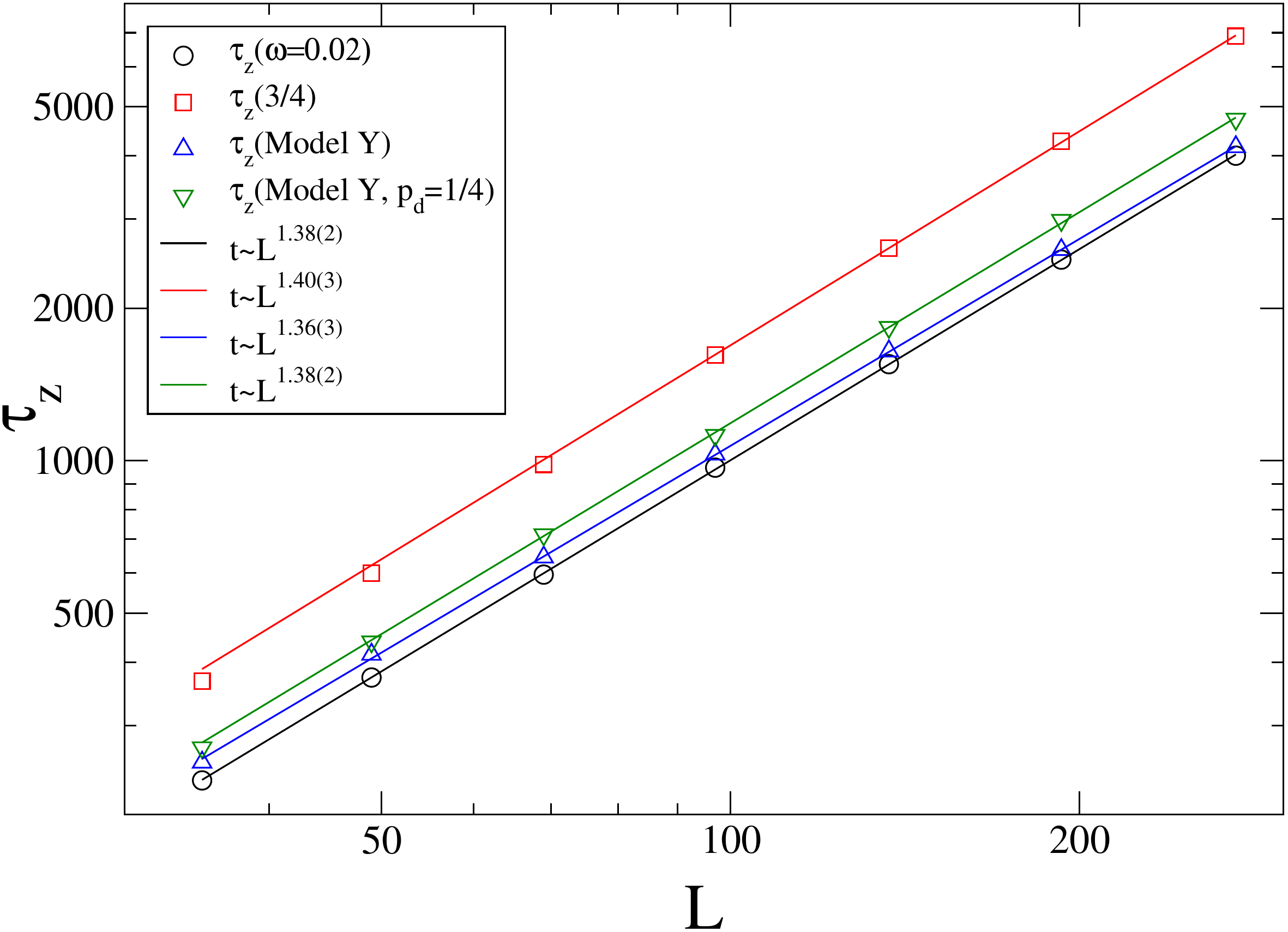}
\caption{Log-log plots of the zipping times as functions of the polymer
length for $\omega=0.02$. The four data sets correspond to four
different choices of the parameters and models of the simulations. Both
for models~B and Y, for different values of the double stranded mobility
($p_d$) and with a different criterion for the definition of the zipping
time we find (the time to have $3L/4$ monomers bound) give a constant
value for $\alpha = 1.37(2)$.
}
\label{taus}
\end{figure}

We investigated the scaling of the zipping time for different values
of the parameters. Figure~\ref{taus} shows the case $\omega = 0.02$ and
compare model~B with the model~Y, using different values for the parameter
$p_d$ (the mobility of the zipped part of the polymer) and using an
alternative definition of zipping time. In the latter we used the time
needed to reach a configuration where $3L/4$ monomers are bound for the
first time.  Apart from a global shift of the time scales we find that the
two criteria of defining the zipping time from $n_b=L/2$ or $n_b=3L/4$
yield the same value of the exponent $\alpha$. In all cases analyzed
in model~B and model~Y the estimated exponent are consistent. Again we
find as final estimate $\alpha = 1.37(2)$. In the zipping dynamics and
for the lattice models studied here this exponent is particularly robust
and is observed in different temperature regimes and at different values
of the parameters.

\begin{figure}[ht!]
\centering
\includegraphics[height=8cm]{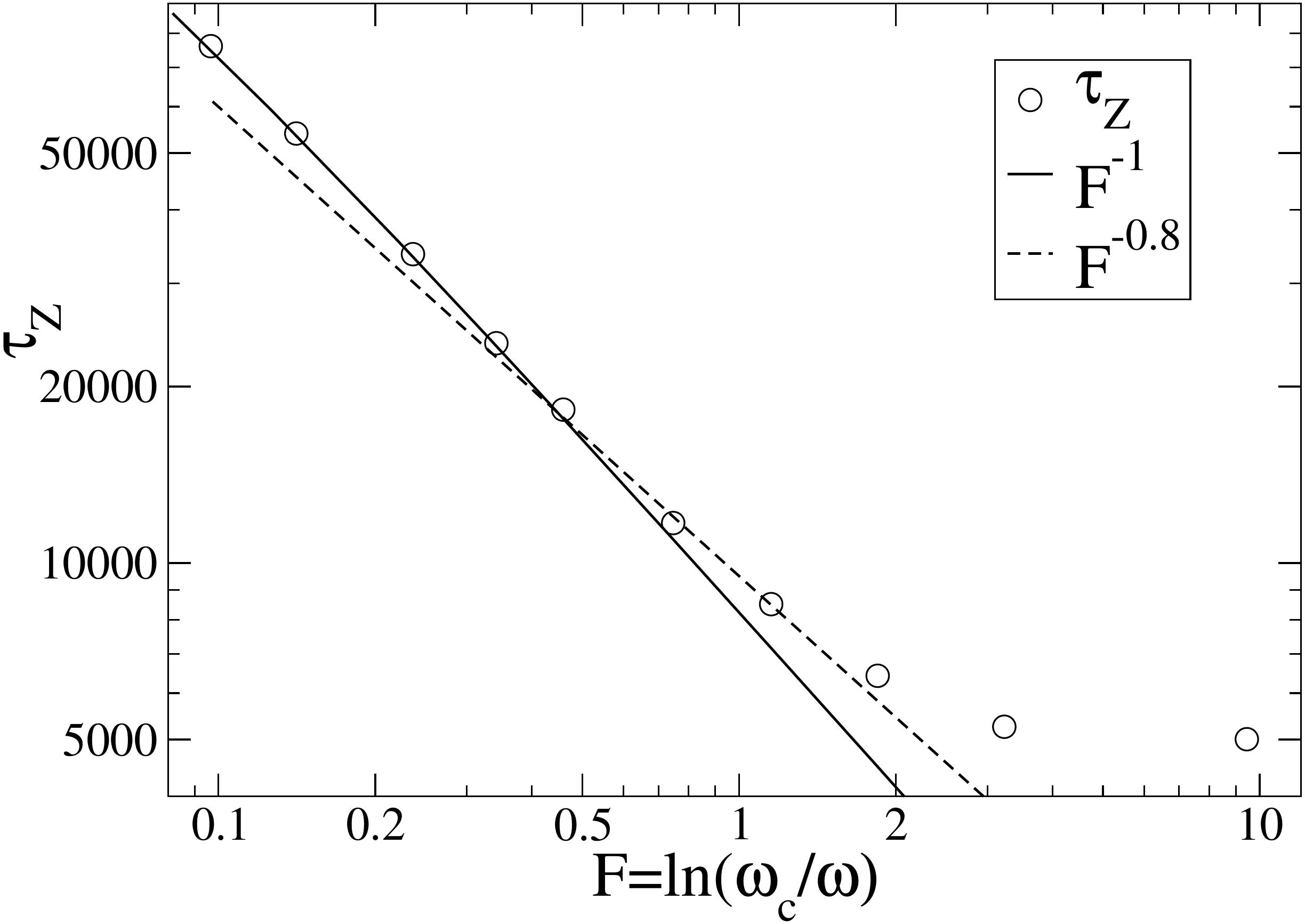}
\caption{Plot of the unzipping time $\tau_z$ as a function of the
parameter $\ln (\omega/\omega_c)$ (the driving force) for $L=385$ and
model~B. Here $\omega_c=\exp(\varepsilon/k_B T_c)$ is the value of the 
weight at the thermal unzipping temperature. At small forces the scaling
is consistent with $\tau_z \sim 1/F$. The dashed line shows a comparison
with the results of Ref.~\cite{luo09}, expected to be valid at stronger
forcing.}
\label{force}
\end{figure}

We turn now to the study of the dependence of the zipping time on
the driving force. In the case of forced polymer translocation,
the driving force is the difference in chemical potential for
monomers on the two sides of the separating membrane. In our model
one can view the equilibrium critical point ($T=T_c$) as the limit
of zero applied force. We define then the driving force as $F = \ln
(\omega_c/\omega)$, where $\omega$ is the Boltzmann weight associated to
the unbinding of two monomers and $\omega_c$ its value at the critical
point. Figure~\ref{force} shows a plot of the zipping time for model~B
with $L=385$, as a function of $F$. Extrapolation of the data at
weak forces ($F \leq 0.5$) yields a scaling $\tau_z \sim F^{-\gamma}$
with $\gamma = 0.92(5)$. This is close to the scaling $\tau_z \sim 1/F$
observed in polymer translocation \cite{vock08}. The small deviation from
the $1/F$ scaling is likely due to the fact that the data are not in the
full asymptotic regime. In addition, the shape of the $\tau_z$ vs. $F$
data in Fig.~\ref{force} shows a turnover from the linear response regime
$\sim 1/F$ towards a smaller slope at stronger forces which is very similar
to the $\tau$ vs. $1/F$ plot observed in translocation \cite{luo09}.
The dashed line in Fig.~\ref{force} has a slope $\gamma =
0.8$, as observed for simulations of polymer translocaltion for for
strong forces \cite{luo09}, beyond the linear regime. Obviously this
intermediate regime is rather narrow and it is difficult to characterize
it from the analysis of the simulation data. Recent analytical work
\cite{saka07,saka10} on a simplified model of polymer translocation
predicts the existence of an intermediate regime where the $\tau_t
\sim 1/F$ breaks down. This is consistent with our findings, however the
exponents~\cite{saka10} do not seem to match the numerical results for the
zipping dynamics.  This is an interesting point which deserves further
theoretical investigations. In our simulations, beyond $F \gtrsim 3$,
the zipping time is weakly dependent on $F$, and the $\tau_z$ vs. $F$
tends towards a flat asymptotic limit at large forces. We stress that
for the whole range of forces shown in Fig.~\ref{force} and up to $F
\to \infty$, the scaling of $\tau_z$ vs. $L$ is governed by an exponent
$\alpha \approx 1.37$.

To gain some more insight on the polymer conformation we investigated
the radius of gyration of the two single strands and that of the double
stranded part during zipping. We restricted ourselves to model~Y in which
inner loops in the zipped strand are suppressed. For polymers of length
$L$ we computed the radius of gyration when each individual Monte Carlo
run reaches $L/2$ bound monomers for the first time. The average value
over different independent runs is then taken.  As no loops are allowed,
the configuration with $L/2$ bound monomers corresponds to that of a star
polymer with three arms of length $L/2$, each.

\begin{figure}[t]
\centering
\includegraphics[height=8.0cm]{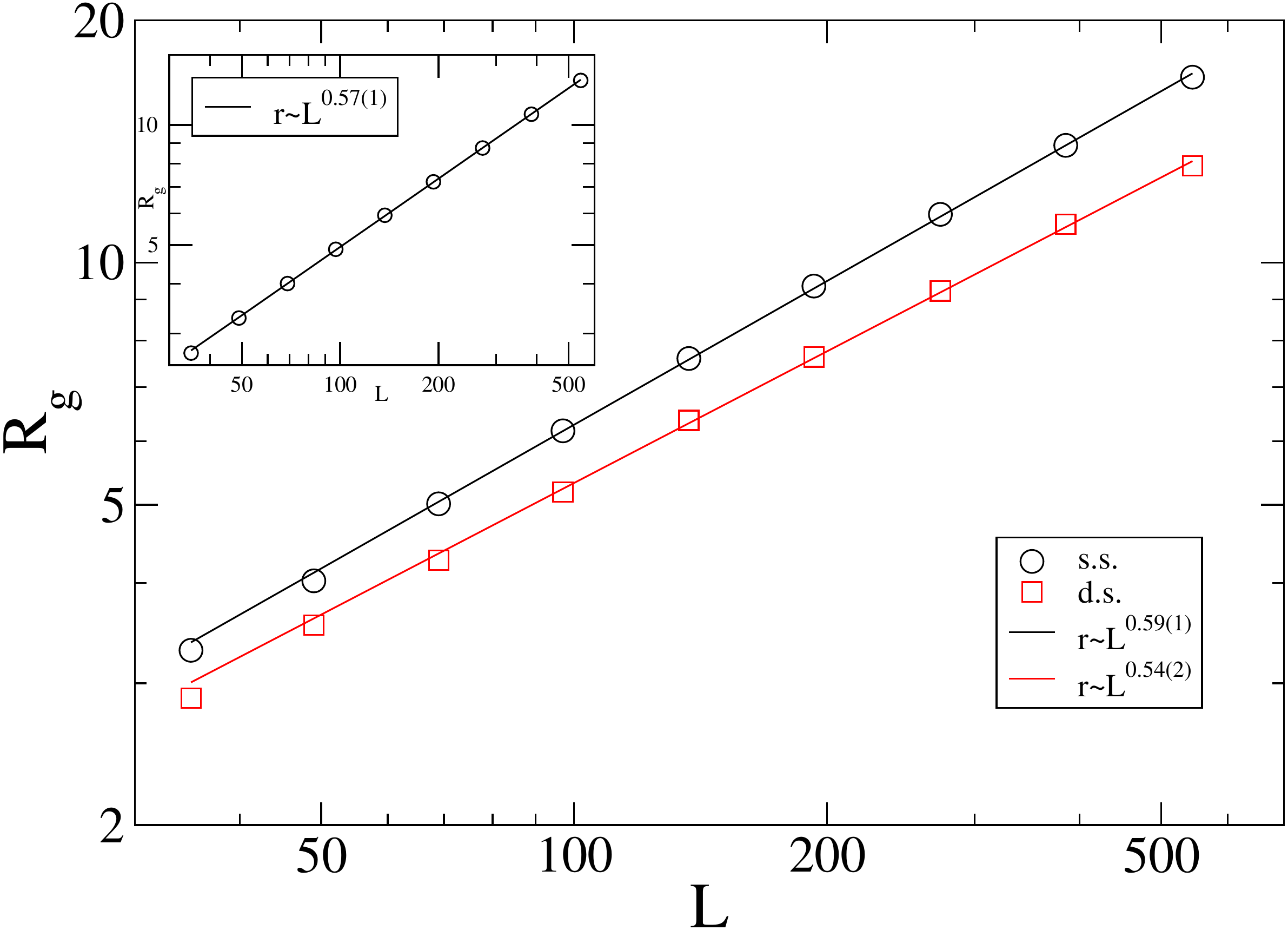}
\caption{Plot of the radius of gyration of the single stranded segments
(circles) and of the zipped part (squares) at zipping for polymers
of different length. Calcualtions have been performed on the final
configuration after zipping of half of the polymer in the model~Y. Here
$\omega=0.02$. In the inset we show a plot of the radius of gyration
vs. $L$ of an equilibrated star polymer with three branches of length
$L/2$.}
\label{gyration}
\end{figure}

Figure~\ref{gyration} plots the radius of gyration of the single
strands and of the zipped part of the polymer as a function of $L$
and in the inset that of an equilibrated star polymer with three arms
of lengths $L/2$, for a comparison. There are some differences in the
scaling of the radius of gyration for the single stranded segments
($R_g \sim L^{0.59(1)}$) compared with that of the zipped part ($R_g
\sim L^{0.54(2)}$). The latter appears to be slightly more compressed
compared to the prediction from the equilibrium scaling $R_g \sim L^\nu$,
with $\nu = 0.588$.  Luo et al.~\cite{luo09} discussed the scaling of
the gyration radius of a translocating polymer in the regime of ``fast"
translocation where the translocation time scales with an exponent
$\alpha \approx 1.37$. They found that for the radius of gyration
after translocation scales as $R_g \sim L^{0.51(1)}$, which is quite
consistent with our estimate $0.54(2)$ for the zipped part. However,
our value does not significantly differ from the equilibrium scaling;
deviations from the Flory exponent could be due to finite size correction.

\begin{figure}[ht!]
\centering
\includegraphics[height=8.0cm]{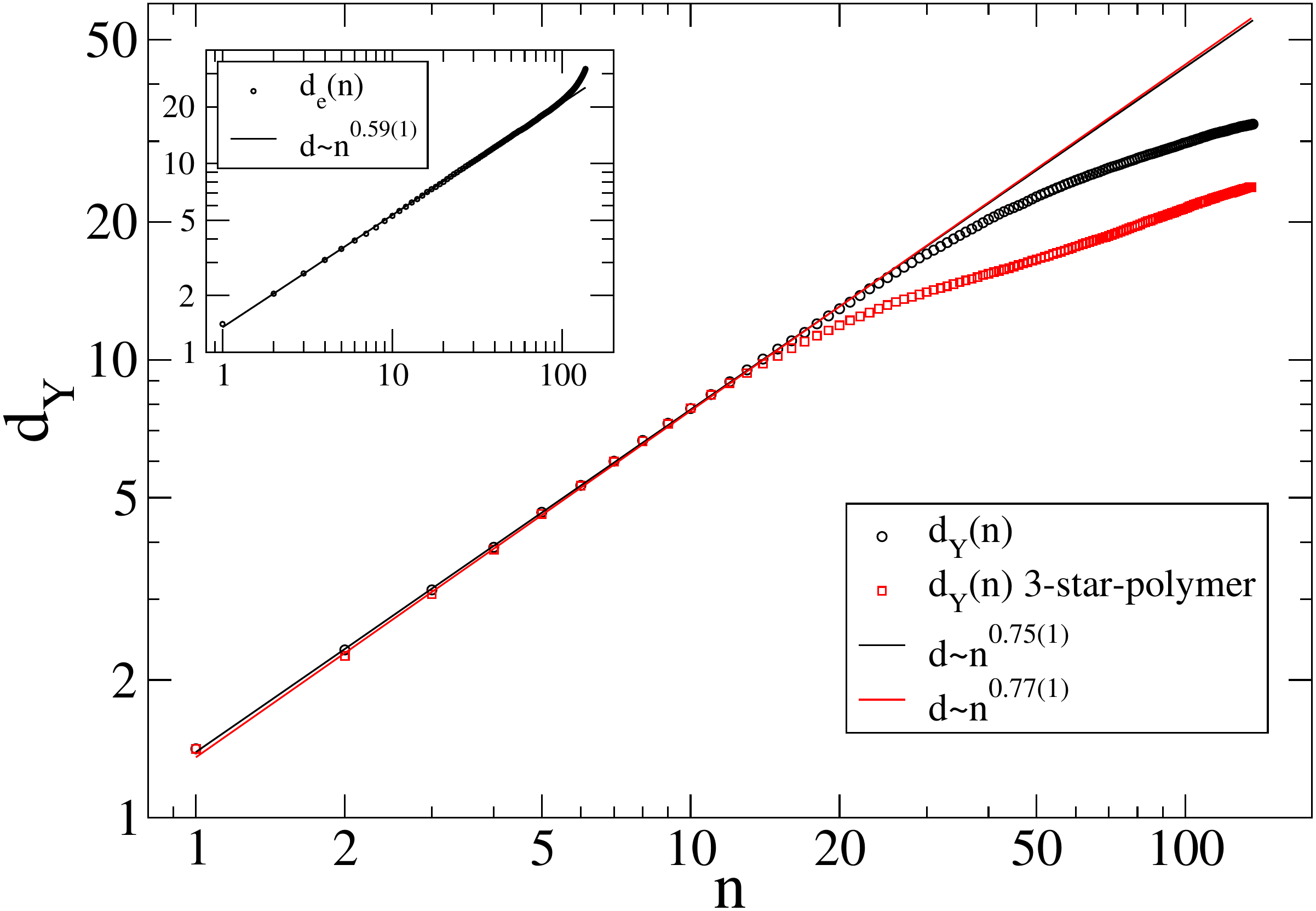}
\caption{Plot of $d_Y(n)$ (Eq.~(\ref{dY})) as a function of $n$. Black
circles refer to the configuration of the two strands right after the
zippping of half of the chains ($L=273$, $\omega=0.02$), while red squares
refer to equilibrium configurations of a star polymer with three branches
of length $L/2$.  Inset: Plot of $d_e(n)$ (Eq.~(\ref{de})) as a function of $n$,
for $L=273$, $\omega=0.02$. The results show that the conformation of the
polymer at the far ends in in equilibrium.}
\label{den}
\end{figure}

The local conformation of the single strands at zipping can be analyzed 
from the scaling of the average distance from the junction 
\begin{equation}
d_Y (n) = \frac 1 2 \sum_{i=1}^2 \left\langle 
\sqrt{ \left( \vec{r}_{L/2}^{\, (i)} 
- \vec{r}_{L/2+n}^{\, (i)}\right)^2} \right\rangle
\label{dY}
\end{equation}
where $\vec{r}_k^{\, (i)}$ denotes the position of the $k$-th monomer
of the strand $i$, the sum is an average on the two strands and
$\langle \rangle$ denotes the average over $10^3$ independent Monte
Carlo runs.  We focus here to the case where $n$ is positive, i.e. to
monomers belonging to the single strands.  This quantity is plotted in
Fig.~\ref{den} for strands of length $L=273$. Again the data are taken
at the time when half of the bases are bound in model~Y.  We find
that $d_Y(n) \sim n^{0.75(1)}$ for $n \lesssim 20$, with deviations
from this scaling behavior for larger $n$, i.e. further away from the
branching point. For a comparison we also plot the same quantity for an
equilibrated star polymer with three arms. For the star polymer at small
$n$ we find $d_Y(n) \sim n^{0.77(1)}$ which is indeed a similar scaling
as the zipping polymer.  However the polymer during zipping is ``more
stretched" than a star polymer at equilibrium as shown in Fig.~\ref{den}.
To our knowledge the scaling behavior in the vicinity of the contact
point for a star polymer has not been investigate yet.

The inset of Fig.~\ref{den} shows a plot of
$d_e (n)$ defined as
\begin{equation}
d_e (n) = \frac 1 2 \sum_{i=1}^2 \left\langle 
\sqrt{ \left( \vec{r}_{L}^{\, (i)} 
- \vec{r}_{L-n}^{\, (i)}\right)^2} \right\rangle,
\label{de}
\end{equation}
which is the average distance measured from the end of the single
strands. For this quantity we find a scaling in good agreement with
$d_e (n) \sim n^\nu$, which implies that the local conformation close
to the ends of the single strands are well-equilibrated. Thus, the
analysis shows that the single strands are in a stem-flower conformation
\cite{broc93,saka07}, where the monomer density in the flower is described
by the Flory exponent $\nu$. The stem is somewhat more stretched than an
equilibrated star polymer, where the stretching is due to self-avoidance
between the three arms which join a common contact point.

\begin{figure}[ht!]
\centering
\includegraphics[height=8.0cm]{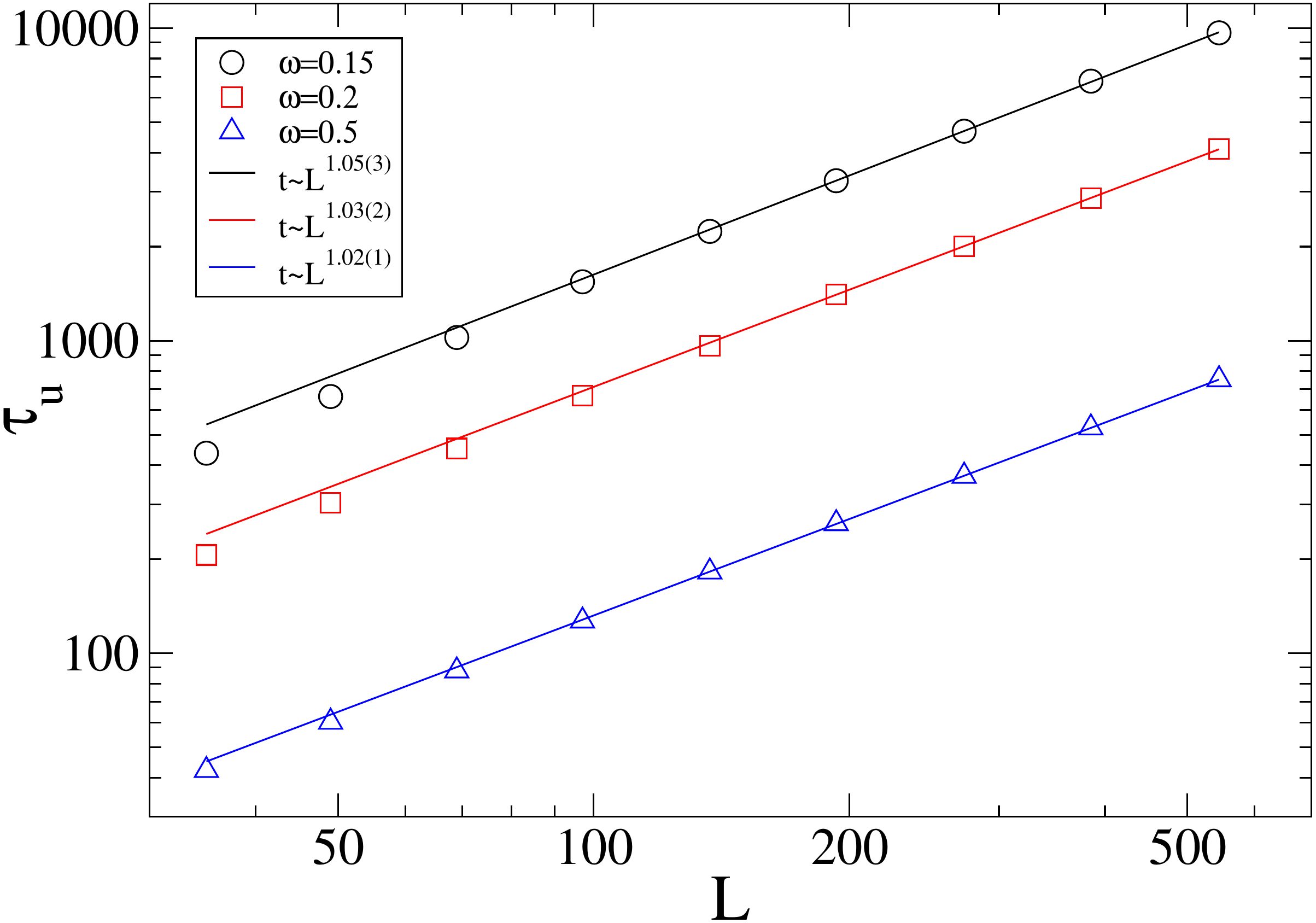}
\caption{Plot of the unzipping time $\tau_u$ as a function of the chain 
length $L$ (model~Y). Different colors stand for different values 
of $\omega$.}
\label{unzip}
\end{figure}

We now turn to the reverse process of unzipping. We start from an
equilibrated double stranded conformation at $T=0$ and then raise the
temperature to a value $T>T_c$. Note that, because of the double strand
move incorporated in our model, a $T=0$ conformation is not frozen. A
move is attempted with rate $p_d$. The temperature raise produces the
unbinding of the two strands. In model~B this is enhanced by the
formation of loops proliferating along the zipped segment. In model~Y
the contact point between the zipped strand and the single stranded
segments performs a biased diffusive motion. For model~Y our simulation results
show that the unzipping time, defined as the average time needed to get
half of the monomers unbound, scales as $\tau_u \sim L$. This is shown in
Fig.~\ref{unzip}, for different values of the parameter $\omega>\omega_c$.
In this case the dynamics is thus not anomalous.

\section{Discussion}

In this paper we studied the zipping dynamics of polymers.  Our numerical
analysis showed that the zipping process is characterized by an anomalous
exponent, with zipping times scaling as $\tau_z \sim L^{1.37(2)}$.
Two different models were considered, with and without inner loops in
the double stranded configurations. In addition we investigated a wide
range of temperatures and parameters. For all the values investigated
the exponent $\alpha \approx 1.37$ was confirmed. The asymptotic behavior
sets in already for short polymers at low temperatures. Numerical results
suggest that the scaling as a function of the driving force is $\tau_z
\sim 1/F$ in the limit $F \to 0$.  The analysis of the gyration radii,
and the comparison with those of an equilibrated star polymer, shows
that the polymer configuration differs somewhat to an equilibrium one: in
the vicinity of the contact point between the strands a zipping polymer
is more stretched than an equilibrium polymer. However the scaling of
the global radii of gyration do not differ significantly from their
equilibrium counterparts.

\begin{figure}[ht!]
\includegraphics[height=6cm]{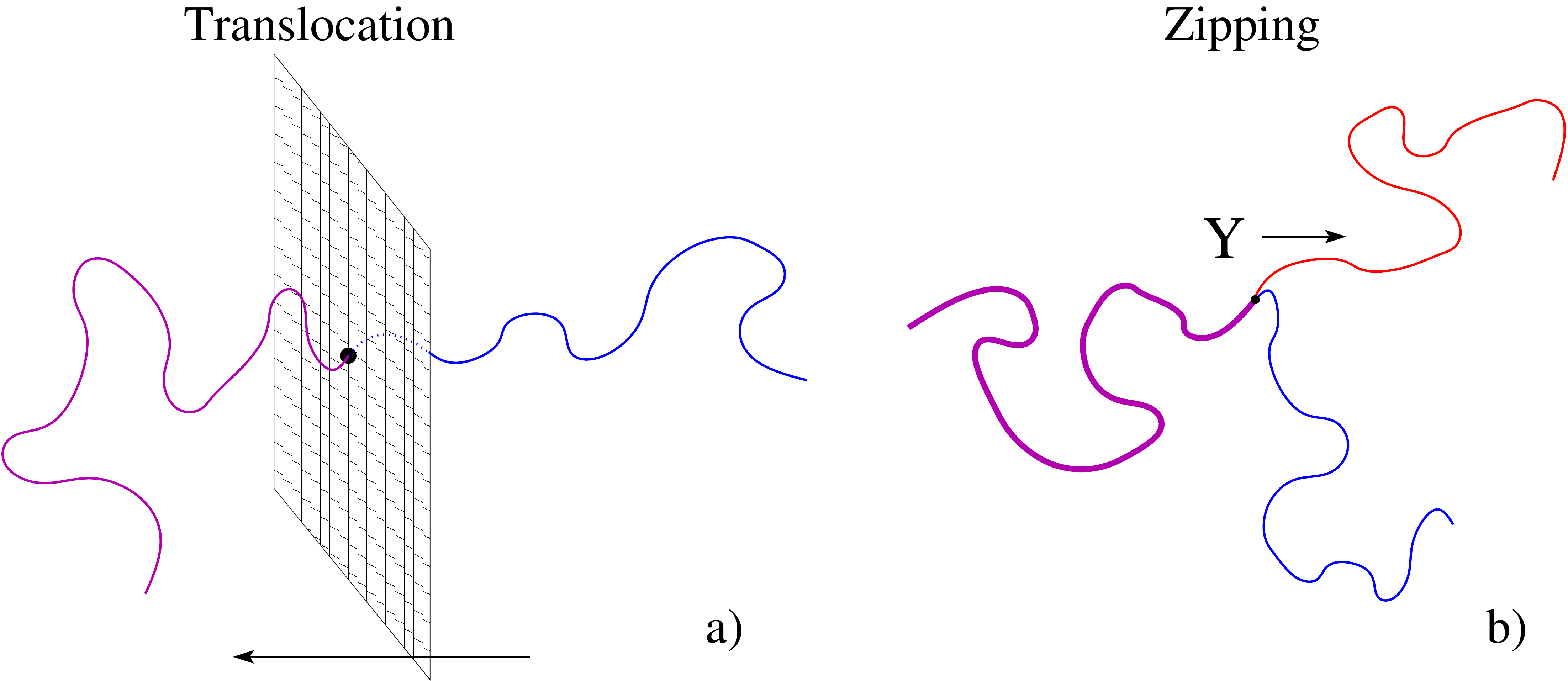}
\caption{Analogies between forced polymer translocation (a) and
polymer zipping (b). Translocation is biased from the right side to
the left side of the separating membrane. As a result the size of the
left portion of the polymer increases with time.  At temperatures below
the unbinding temperatures, the zipped configuration (thick strands)
is favored over the unzipped one (thin polymer strands). As a result the
size of the zipped part increases in time. Both processes are described by
similar free energies, which for the leading orders are for translocation
$F(s) = \mu_+ s + \mu_- (L-s)$ ($s$ is the length of the side of the
polymer, $\mu_+$ and $\mu_-$ the chemical potentials at the two sides),
while for zipping $F(s) = f_z s + 2 f_u (L-s)$ ($s$ is the length of the
zipped strand, $f_z$ and $f_u$ the free energies per unit of length of
the zipped and unzipped strands). In both cases the dynamics is anomalous.}
\label{transzip}
\end{figure}

We note that the exponent found in our simulations is in good agreement
with that observed in forced translocation \cite{vock08}, where a chemical
potential difference drives a polymer to cross a membrane through a narrow
pore. Also polymer absorption seems to be governed by the same type of
exponent, at least for weak adsorption energies~\cite{panj09}.  In the
zipping process discussed in this paper the polymer does not cross any
planar surface and it is driven by monomer-monomer binding to close up and
form a double stranded conformation. Apart from geometrical differences
there are strong analogies (see Fig.~\ref{transzip}) between zipping
and translocation and it is very likely indeed that they share the same
type of anomalous dynamics. If this is the case, the exponent $\alpha$
governing the anomalous scaling is not influenced by the geometry of
the problem, i.e. whether there is a plane (as in translocation and
adsorption) or a three-polymer contact point (in zipping).

A recent theory for translocation \cite{vock08} predicts $\alpha =
(1+2\nu)/(1+\nu) \approx 1.37$ and $\tau_t \sim 1/F$. This theory is
valid at weak forcing as it assumes that the polymer conformation is
not different from that in absence of driving. It is based on memory
effects, which arise due to the presence of a chain tension imbalance
in the vicinity of the pore in the case of translocation. In the case
of zipping, we indeed observed some chain tension in the vicinity of the
contact point, which causes a stretched ``stem" configuration discussed
above. In the case of unzipping no tension is present. The excess of
monomers leading to a non-equilibrium conformation during unzipping
is dissipated in other directions and does not influence the contact
point dynamics. This supports indeed the idea that imbalance in the
chain tension is the origin of the anomalous exponent. At stronger
forcing we find that the zipping time scales non-linearly as
a function of $F^{-1}$, in a similar way as found in simulations of forced 
translocation \cite{luo09}. To our knowledge this intermediate regime 
is still poorly understood. Recent theoretical work \cite{saka10} highlighted
the complexity of the problem and predicted the existence of various translocation
regimes. It is still unclear if this scenario is also valid for the zipping process.

Finally, systems where the effects discussed in this paper could be
experimentally investigated are DNA or RNA hairpins.  These are single
stranded molecules which are composed by two self-complementary halfs.
The self-complementarity drives the formation of a double-helical
conformation at low temperatures, eventually terminating with a loop.
Various aspects of the kinetics of DNA hairpins formation have been
investigated (see e.g. \cite{bonn98,kim06}), but, to our knowledge,
not the length dependence of the opening/closing dynamics.  Our results
suggest that the scaling as a function of the sequence length of the
zipping and unzipping times would be different, where only the first
of them would be governed by anomalous dynamics.  This is of course
in absence of hydrodynamic interactions, and neglecting the winding
dynamics. The latter may slow down the dynamics even further, as shown
in a recent work \cite{baie10a}.

\section*{Acknowledgements:}

We acknowledge M. Baiesi, G.T. Barkema and D. Panja for useful
discussions.  We acknowledge financial support from Research Foundation
Flanders, Fonds voor Wetenschappelijk Onderzoek (FWO), grant G.0311.08

\end{document}